\documentclass[draft]{agujournal2019}
\usepackage{graphicx}
\usepackage{url} 
\usepackage[inline]{trackchanges} 
\usepackage{soul}
\usepackage{subfigure}
\usepackage{multirow}
\usepackage{ragged2e}
\justifying
\usepackage{mathtools}
\newcommand{\defeq}{\vcentcolon=}

\draftfalse

\journalname{Space Weather}
\begin{document}

\title{MEMPSEP I : Forecasting the Probability of Solar Energetic Particle Event Occurrence  using a Multivariate  Ensemble of Convolutional Neural  Networks}

\authors{Subhamoy Chatterjee\affil{1}, Maher Dayeh\affil{2,3}, Andr\'{e}s Mu\~{n}oz-Jaramillo\affil{1}, Hazel M. Bain\affil{4,5},  Kimberly Moreland\affil{2,3,4,5},  Samuel Hart\affil{2,3}}

\affiliation{1}{Southwest Research Institute, Boulder, CO, USA}
\affiliation{2}{Southwest Research Institute, San Antonio, TX, USA}
\affiliation{3}{The University of Texas at San Antonio, San Antonio, TX, USA}
\affiliation{4}{Cooperative Institute for Research in Environmental Sciences, University of Boulder, CO, USA}
\affiliation{5}{Space Weather Prediction Center, NOAA, Boulder, CO, USA}

\correspondingauthor{Subhamoy Chatterjee}{subhamoy.chatterjee@swri.org}

\begin{keypoints}
\item End-to-End Deep Neural Network model-ensemble trained on remote sensing + in-situ dataset for forecast of SEP occurrence

\item  Usage of model ensemble maximizes the use of an imbalanced dataset and brings forecasting confidence with uncertainty estimates

\item Calibration of model outcome to true probability optimizes the forecast reliability acting as a means to step away from the binary forecast

\end{keypoints}

\begin{abstract}
The Sun continuously affects the interplanetary environment through a host of interconnected and dynamic physical processes. Solar flares, Coronal Mass Ejections (CMEs), and Solar Energetic Particles (SEPs) are among the key drivers of space weather in the near-Earth environment and beyond. While some CMEs and flares are associated with intense SEPs, some show little to no SEP association. To date, robust long-term (hours-days) forecasting of SEP occurrence and associated properties (e.g., onset, peak intensities) does not effectively exist and the search for such development continues. Through an Operations-2-Research support, we developed a self-contained model that utilizes a comprehensive dataset and provides a probabilistic forecast for SEP event occurrence and its properties. The model is named Multivariate Ensemble of Models for Probabilistic Forecast of Solar Energetic Particles (MEMPSEP). MEMPSEP workhorse is an ensemble of Convolutional Neural Networks that ingests a comprehensive dataset (MEMPSEP III - \cite{moreland2023}) of full-disc magnetogram-sequences and in-situ data from different sources to forecast the occurrence (MEMPSEP I - this work) and properties (MEMPSEP II - \citeA{dayeh2023}) of a SEP event. This work focuses on estimating true SEP occurrence probabilities achieving a 2.5\% improvement in reliability and a Brier score of 0.14.   The outcome provides flexibility for the end-users to determine their own acceptable level of risk, rather than imposing a detection threshold that optimizes an arbitrary binary classification metric.   Furthermore, the model-ensemble, trained to utilize the large class-imbalance between events and non-events, provides a clear measure of uncertainty in our forecast.

\end{abstract}

\section{Introduction}

The interplanetary (IP) environment is constantly influenced by the Sun through a multitude of interconnected and dynamic physical processes collectively known as space weather. Space weather is characterized by phenomena such as solar flares, coronal mass ejections (CMEs), and solar energetic particles (SEPs), which are significant drivers of space weather both near Earth and in distant regions. While some CMEs \cite{Gopalswamy2006, Webb2012, Dayeh2015, Kilpua2019} and flares are associated with large enhancements of SEPs \cite{Gopalswamy2008,Cliver2018}, others display minimal or no correlation with these energetic particles \cite{Marque2006,Swalwell2017,Gopalswamy2018}. Furthermore, establishing a consistent link between the properties of SEPs observed at 1 au, their arrival times, and their origins near or around the Sun is often challenging due to the highly complex nature of the environment that governs the origin, acceleration, and transportation of SEPs in interplanetary space (\citeA{Dayeh2010}, \citeA{Kozarev2010}, \citeA{WHITMAN2022} and references therein).

High energy particles associated with SEP events can impact human health and technologies in a number of ways. The enhanced radiation environment can pose a health hazard for astronauts in space as well as passengers and crew on polar and/or high-altitude flight routes. SEPs can cause single event effects (SEEs), internal charging (IC), and electrostatic discharge (ESD) in satellite electronics when the charge is deposited in circuitry, leading to satellite malfunction \cite{Jiggens2014, Dyer2020, Kottaras2022}. Surface charging (SC) can produce high voltages, leading to arcs and electromagnetic interference, and long-term exposure to ionizing radiation can cause further damage. Protons associated with SEPs can also disrupt and degrade high-frequency (HF) radio communications in polar regions. 

Forecasting SEP events with significant lead time remains a challenge. Robust long-term (hours-days) forecasting of SEP occurrence and associated properties (e.g., onset time, peak particle intensities, heavy ion composition) does not effectively exist and the search for such development continues. The NOAA Space Weather Prediction Center (SWPC) uses the Solar Radiation Storm ``S-scale" to communicate the severity of $\ge$ 10 MeV proton events as observed by the GOES spacecraft. An S1 proton event corresponds to the period of time when the GOES observes $\ge$ 10 MeV protons exceeding 10 particle flux units (1 p.f.u. $\defeq$ 1 particle / ($\mathrm{cm^{-2}}$ s sr)). S2 through S5 proton events correspond to order of magnitude increases in the proton flux i.e. the S2 threshold is 100 p.f.u, the S3 threshold is 1000 p.f.u. and so on. SWPC provides probabilistic proton event forecasts for S1 storms 1, 2, and 3 days into the future. A record of these forecasts can be found at NOAA National Centers for Environmental Information (NCEI) Space Weather Products archive under Daily Reports (https://www.ngdc.noaa.gov/stp/spaceweather.html). For an in-depth discussion of SWPC’s proton forecasts, how they are generated and SWPC’s performance metrics and forecast skill during Solar Cycles 23 (SC23) and 24 (SC24) the reader is directed to \citeA{bain21}. However, there are a few key takeaways from this forecast verification study. “\emph{SWPC probabilistic forecasts struggle to accurately forecast the onset of S1 storms ahead of time. However, in general, once the threshold has been crossed SWPC day 1 forecast probabilities are reliable.}” And “\emph{Reliability diagrams…revealed a tendency to over issue forecast probabilities in the $<$40\% range for day 1 forecasts during Solar Cycle 24. Similarly, there is a tendency to over forecast in the $\ge$80\% probability range for day 3 forecasts in Solar Cycle 24.}” 

The \citeA{bain21} verification study highlights the well-recognized difficulties in forecasting when an active region will erupt and whether an associated flare or CME will accelerate particles which are then observed as an SEP event at Earth. As such, SWPC 1, 2, and 3 probabilistic forecasts are supplemented with short-term (minutes-hours) deterministic Warning and Alert products which communicate to customers when a forecaster believes an event is imminent and when an event onset has been observed, respectively. Weekly Reports of SWPC Warning and Alerts are also archived by NOAA NCEI Periodic Reports. Verification of SWPC Warning product performance metrics during SC23 and SC24 are detailed in \citeA{bain21}. 

A recent review article by \citeA{WHITMAN2022} provided a comprehensive overview of the large variety of current SEP forecasting models encompassing physics-based, empirical, and machine learning (ML) models and that the outputs may be categorical, binary, probabilistic, or deterministic. The article highlights the potential of ML models in ingesting multidimensional data, providing faster inference as compared to physics-based models, and discovering relationships not possible by simple models, and also points out the challenges of using ML models, especially in situations such as data imbalance. 

The complex nature of SEPs motivates us to build a multivariate model that can ingest remote sensing inputs such as solar images and in-situ parameters related to solar wind properties, interplanetary magnetic fields, etc. Convolutional neural networks (CNNs) are efficient in ingesting such complex sets of inputs. It should be noted that probabilistic forecast models are more desirable in the community than the ones that only generate binary (event/non-event) outcomes \cite{bain21}. There are SEP forecasting models that generate probabilities of SEP occurrence at multiple integral energy bands through simple techniques (see \citeA{WHITMAN2022} and references therein). For example \citeA{Papaioannou2022} uses the Bayes approach to estimate SEP occurrence probability by ingesting probability distributions of associated scalar parameters such as CME angular widths, speed, and solar flare longitude, peak strength. However, complex models such as CNNs often suffer from low reliability in forecasting the probability of event occurrence. Reliable forecasts require a good match between event probability and frequency. Also, relying on the outcome of a single model can become detrimental in scenarios where the model is fed with an input that is not well represented by the training set. It has been found that deep learning models can produce confident and wrong outcomes in such situations \cite{duerr2020probabilistic, Chatterjee2022}. Thus, having an ensemble of models can help in understanding the uncertainty in prediction and save us from making incorrect decisions in critical situations. It is also immensely important from the perspective of SEP prediction as such events are much rarer than non-events (e.g. flares or CMEs that do not show a rise in particle flux at Earth above a threshold) and thus there is always a high chance that the training set does not contain all possible conditions associated with an event.

The model, named ``Multivariate Ensemble of Models for Probabilistic Forecast of Solar Energetic Particles" or MEMPSEP, is detailed in three papers. This work describes the ensemble modeling towards a forecast of the true probability of SEP occurrence, \citeA{moreland2023} describe the database, and \citeA{dayeh2023} describe the regression model which enables the forecasting of SEP properties at 1 au.
In particular, we discuss here how we mitigate the problem of class imbalance, build an ensemble of CNNs, and calibrate the CNN outcomes to represent the actual probability of SEP event occurrence with associated uncertainty. Section 2 briefly describes the dataset, section 3 illustrates the ground-truth assignment, section 4 depicts the forecasting model architecture, section 5 illustrates the data management process to generate a model ensemble, section 6 depicts the importance of different model inputs to improve forecast quality, section 7 describes the design of a test set to estimate the model performance,  section 8 shows the calibration of neural network outcome to reliable forecasts, section 9 describes different performance metrics to identify the strengths and weaknesses of the model, section 10 illustrates the model-ensemble performance on the test set through different metrics, section 11 depicts model validation on SHINE 2022 challenge events and non-events, and we finally discuss our results and conclude through section 12.

\section{Pre-Eruptive Data Description} 
The full dataset used in MEMPSEP development and training comprises remote and in situ parameters, both measured and derived. The full dataset is described in detail in MEMPSEP-III. The following subsections briefly describe some of the important data inputs. 
\subsection{Upstream Properties}
We gather upstream solar wind properties from in-situ measurements using data collected over 6 hours prior to the flare onset. The parameters used for our current study are namely solar wind (SW) temperature, SW velocity, SW density, interplanetary magnetic field (IMF) field strength (B), IMF Bx, By and Bz components. We also utilize suprathermal particle fluxes \cite{Dayeh2009,Dayeh2017} and elemental abundance ratios of heavy ions (e.g., Fe/O, H, O, and Fe). More details on the motivation and calculation of these parameters are provided in MEMPSEP-III. Any missing data in upstream properties is filled in by random selection from distributions corresponding to SEP events and non-events (see section 3 for the definition of events and non-events).

\subsection{Magnetograms}
\begin{figure}[t!]
\hspace{-0.1\textwidth}
\includegraphics[width=1.2\linewidth]{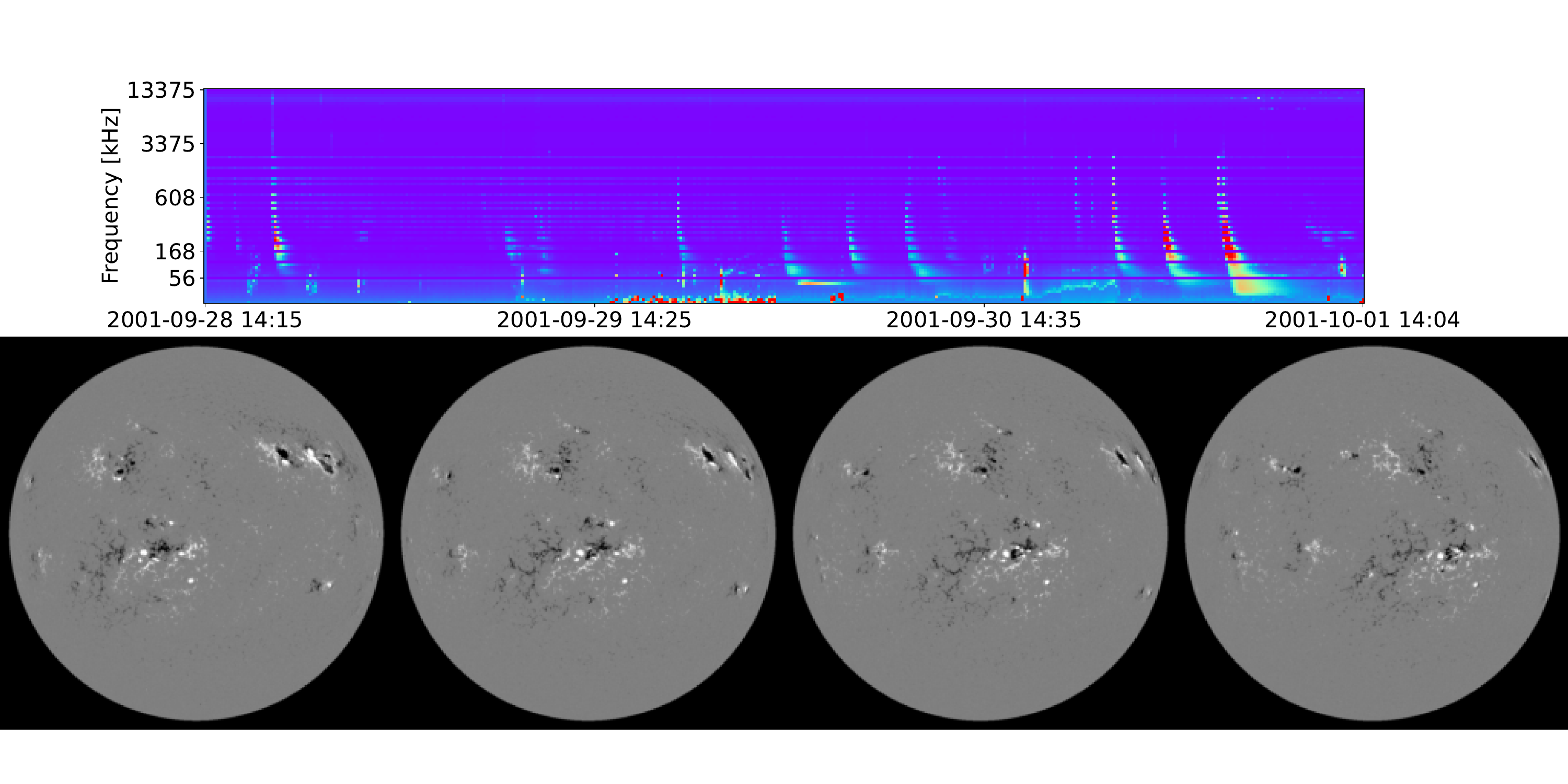}
\vspace{-0.05\textwidth}
\caption{Image data as input to our model prior to flare onset. The upper panel shows the Wind-Waves time-frequency diagram. Every third image in the Sequence of full-disc magnetograms, as input to the CNN, is shown in the lower part.}
\label{fig:mag}
\end{figure}

We use sequences of full-disc line-of-sight magnetograms from the Michelson Doppler Imager onboard Solar and Heliospheric Observatory (SoHO/MDI) \cite{MDI} and the Helioseismic and Magnetic Imager onboard Solar Dynamics Observatory (SDO/HMI) \cite{HMI} as an input to our model. Keeping in mind the memory limitation of GPU we degrade the resolution of magnetograms to 256 pixels $\times$ 256 pixels and collect one magnetogram every 6 hours over 3 days prior to a flare onset. The selection of this temporal span is made to capture the dynamics of the active regions as those rotate through the view. This creates a data-cube of size $256 \times 256 \times 13$. It must be noted that the original size ($1024\times1024$ for MDI and $4096\times4096$ for HMI) and cadence (96 minutes for MDI and 45 seconds for HMI) are degraded here to reduce the memory burden on the computer without losing prior information active region evolution from full-disc magnetograms. As CNNs are blind to the time stamps of input magnetograms, we ensure every two consecutive magnetograms are separated by 6 hours by performing nearest neighbor interpolation for each pixel over the time axis. Nearest neighbor interpolation gets rid of the feature smearing caused by linear interpolation. To homogenize MDI and HMI magnetograms we use a conversion factor of 1.3 (MDI = 1.3*HMI) and clip the field strength within [-1000 Gauss, 1000 Gauss]. This creates a continuous dataset with a longer observational baseline than from a single instrument.  Finally, we normalize the field strength ($f$) using
the transformation $\frac{1}{2}(1+\frac{f}{1000})$. Clipping and normalization are done so that the CNN model does not get driven by outlier pixels. So, pixel values of 0, 0.5 and 1 in the normalized magnetograms (Figure~\ref{fig:mag}) respectively represent -1000 Gauss (black), 0 Gauss (gray),  and 1000 Gauss (white).

\subsection{Wind/Waves time-frequency data}

Studies show SEP correlations with radio burst signatures  \cite{Laurenza09, richardson18} i.e. Type II radio bursts associated with particle acceleration in CME shocks, and Type III radio bursts associated accelerated electrons leaving the Sun along open field magnetic field. We use radio data recorded by the WAVES instrument Radio Receiver Bands 1 and 2 (RAD1: 20 kHz - 1,040 kHz, RAD2: 1.075 MHz - 13.825 MHz) on-board the Wind spacecraft \cite{bougeret95} to identify radio burst signatures indicative of accelerated particles. Of the 256 RAD1 frequency channels, only 20 channels contain observed data, while the others contain interpolated data. For input to the model, we use only RAD1 channels with observed data. All 256 RAD2 channels contained observed data, however, to reduce our data input to the model we use a total of 80 frequency channels combining RAD1 and RAD2.
We down-sample the time axis to a cadence of 10 minutes with an anti-aliasing approach (low-pass filtering + re-sampling), take a logarithm of the intensities and clip them within the range [-1,1]. The logarithmic transformation and clipping of pixel values are done so that the CNN model does not get driven by outlier pixels and instead focuses on the range of values that are more prevalent. Three days of WAVES data prior to a flare onset is used as an input to the model, resulting in a $432\time80$ time-frequency radio image (Figure~\ref{fig:mag}). 

\subsection{X-ray and Electron time-series data}
We use X-ray data from GOES at two channels with 1-minute cadence for 24 hours prior to flare onset. We also use time series of electrons at L1 over 7 channels prior for 24 hours prior to flare onset with a cadence of 10 seconds. More details on the motivation and source of these time series data are provided in MEMPSEP-III.

\section{Assigning Ground-truth}
For our study, ground-truth labels (1: event, 0: non-event) are assigned based on the occurrence or non-occurrence of a proton event at Earth in association with a flare. While it is well understood that large proton events are generally associated with particles accelerated in CME-driven shock fronts, the first indication of an eruptive event available to a forecasters is coronal imagery showing the flare. Currently, images from e.g. the SOHO Large Angle and Spectrometric COronagraph (LASCO; \citeA{LASCO}) which images CMEs leaving the Sun, are not available in real-time. LASCO images can have a latency in excess of 4 to 6 hours, the latency of which is too slow to determine the presence of a CME and make a forecast before the onset of many proton events. For real-time forecasting, we want to train a model which can be issuing forecasts at the first sign of a potential proton event. To identify our event class, we look for integrated ($\ge$ 10 MeV) GOES particle flux crossing a threshold of 5 p.f.u. within a 6 hour period from flare onset. If the particle flux rises above that threshold and the maximum 3-hour pre-flare background flux, a ground-truth of 1 is assigned to the flare, otherwise 0 is assigned. More details about the approach are provided in MEMPSEP-III. Note we use a threshold of 5 p.f.u., which is lower than the SWPC S1 threshold of 10 p.f.u., to increase the number of events with which to train and test the model.

\section{Model description}
\begin{figure}[t!]
\hspace{-0.12\textwidth}
\includegraphics[width=1.2\linewidth]{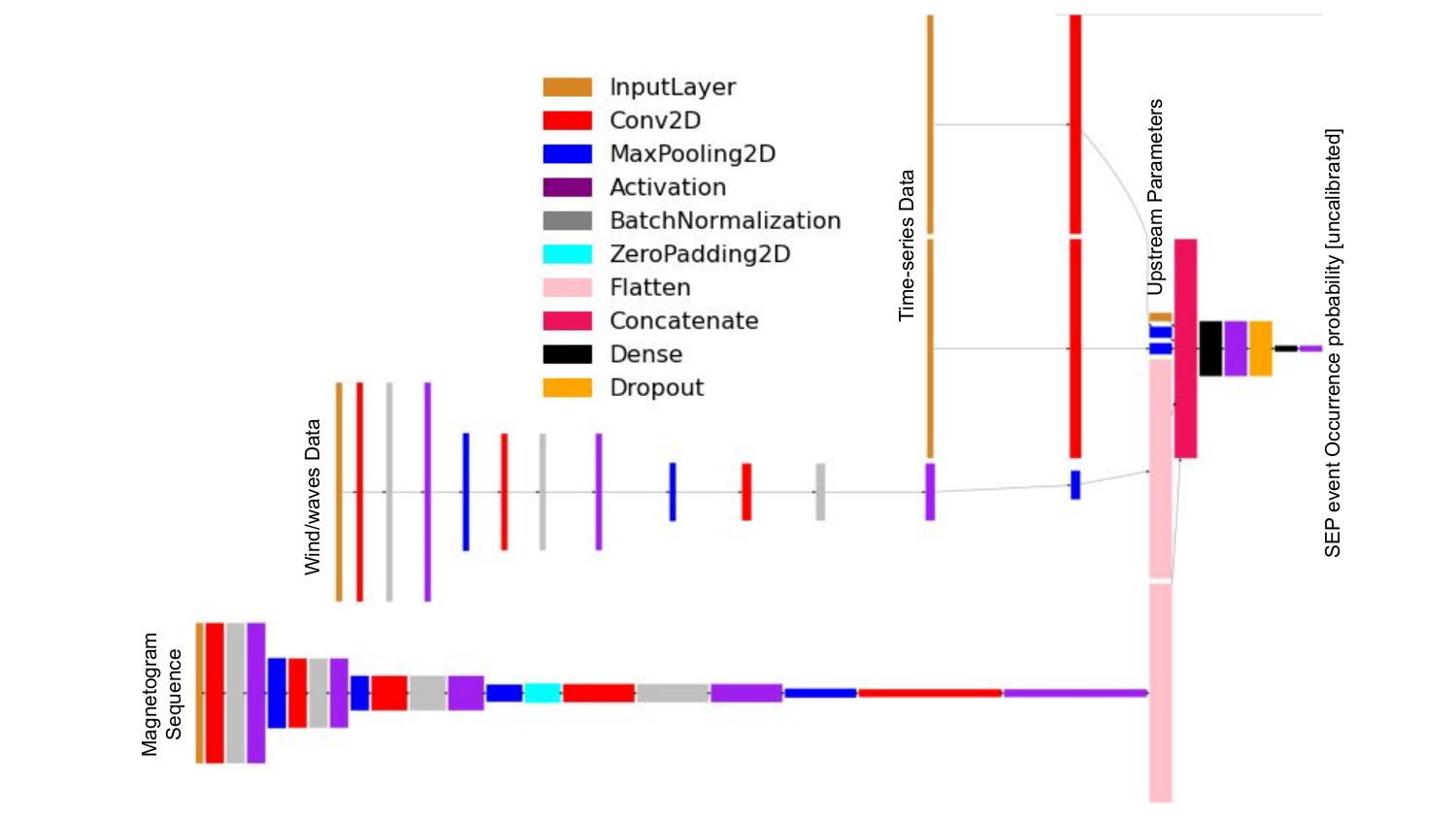}
\vspace{-0.05\textwidth}
\caption{Our CNN model architecture with multiple inputs to predict SEP occurrence probability. The CNN processes the input images with repeated 2D convolution, batch normalization, non-linear activation, and max-pooling. The flattened layers are concatenated and passed through dense layers to produce the SEP occurrence probability (uncalibrated).  }
\label{fig:model}
\end{figure}

We build a Convolutional Neural Network (CNN; \citeA{lecun2015deep}) architecture that ingests multiple inputs such as images, scalar parameters, and time series (Figure~\ref{fig:model}) and predicts the probability of SEP occurrence. The CNN processes the input images image sequences with repeated 2D convolution, batch-normalization, non-linear activation, and max-pooling. The flattened layers are concatenated with scalar parameters and features extracted from time series data, and passed through dense (fully connected) layers to produce the uncalibrated SEP occurrence probability. Dropout is used before the last layer causing random disconnections between nodes and helping in the regularization of the model.  It must be noted that the temporal information the time series data is fused in the initial layers of the CNN and then broken into multiple feature layers that are further passed onto subsequent layers. 
\\ 
To optimize the model's weights and biases we minimize Binary Cross Entropy as a loss function depicted by:
\begin{linenomath*}
    \begin{equation}
        Loss = -\sum_{i=1}^{N}[t_i\log(p_i) + (1-t_i)\log(1-p_i)]
    \end{equation}
\end{linenomath*}

 where $t_i$, $p_i$ respectively represent target class $\in\{0,1\}$ and predicted SEP occurrence probability for training instance $i$. It must be noted that minimizing this loss function drives the model outcome either towards 0 or 1.

\begin{figure}[t!]
\hspace{-0.15\textwidth}
\includegraphics[width=1.3\linewidth]{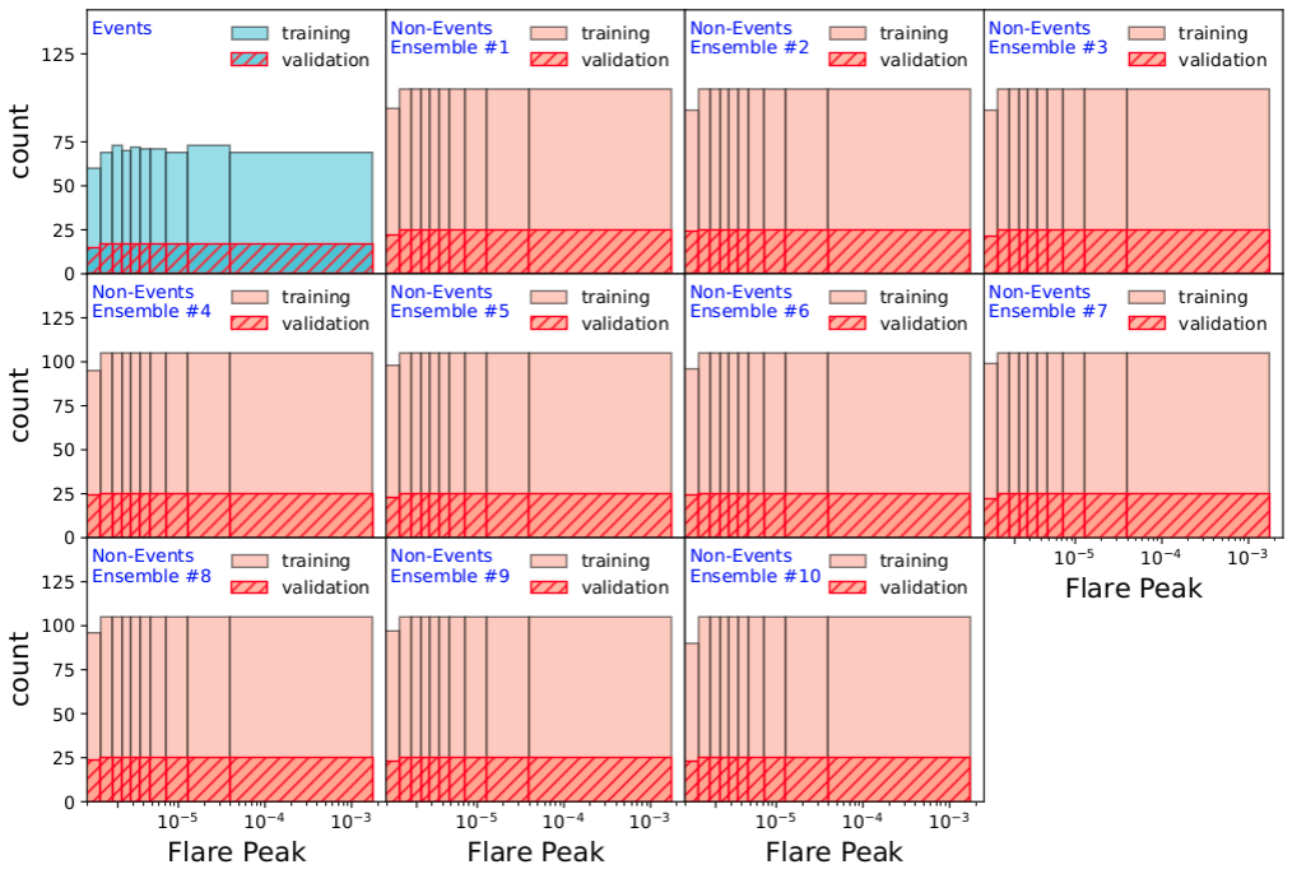}
\vspace{-0.05\textwidth}
\caption{Generation of multiple training and validation sets towards training model-ensemble. Different panels represent equal frequency histograms over flare peaks for training and validation sets. Event histograms are depicted in sky blue and non-event histograms are depicted in pink. It should be noted that for both events and non-events, the validation set contains $\frac{1}{4}^{th}$ the number of data points of those in the training set. Also, all ensemble members have a common set of events and are optimized to have the least overlap for non-events. The total number of non-events in each ensemble member is 1.5 times that of events.}
\label{fig:ensemble}
\end{figure}

\section{Ensemble}
A large class imbalance of SEP events and non-events poses difficulty in the success of Machine learning-based prediction models \cite{kasapis_2022}. We make use of the class imbalance creating multiple training and validation sets combining minimally overlapping non-events and the same set of events.
To realize this, we first evaluate an equal-frequency histogram of flare peak with 10 bins for events. We then use those bins edges to randomly choose non-events (1.5 times the number of events) per bin 10 times with minimal intersection. To ensure minimal intersection we use a membership threshold of $\lceil\frac{mn}{L}\rceil$ for each non-event datapoint given a flare bin where $n$, $m$ and $L$ stand for the number of ensemble members, number of datapoints per selection and the total number of non-events per flare bin respectively. We have about 850 (85 per bin) events and thus for each random selection, we choose about 1275 ($\sim$ 127 per bin) non-events. Thus for 10 random selection of non-event subsets, we utilize close to 10000 non-events. Non-events for each random selection in combination with 850 events are utilized to train and validate a model. We thus train 10 such models creating a model ensemble. Training a single model would have otherwise caused a class imbalance of $\sim$ 1:15 between events and non-events. The data preparation for the training of the model-ensemble is elaborated in Figure~\ref{fig:ensemble}.
We use a 4:1 training:validation split (1020 (non-events) + 680 (events) for training and 255 (non-events) + 170 (events) for validation; see Figure~\ref{fig:ensemble}) and stop training the model when the validation accuracy stops improving.

\begin{figure}[t!]
\hspace{-0.05\textwidth}
\includegraphics[width=1.1\linewidth]{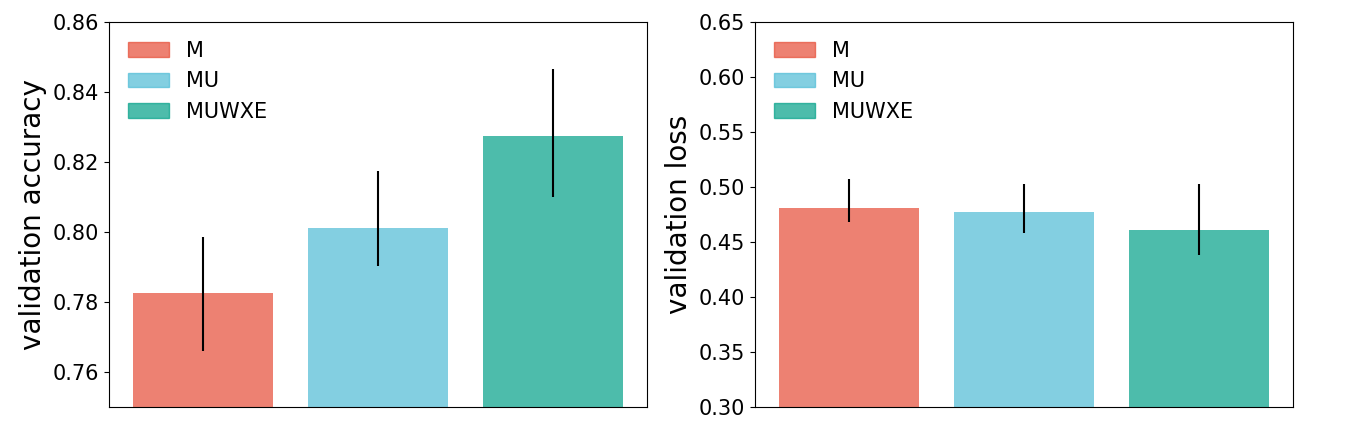}
\vspace{-0.05\textwidth}
\caption{Depiction of the importance of model inputs. The model inputs are masked one by one to understand the change in SEP classification accuracy (defined in section 9.2) on the validation sets. Starting from only Magnetogram sequences (M) it appears that the inclusion of every input such as upstream properties (U) and Wind/Waves time-frequency image (W), X-ray (X) and Electron (E) time series causes improvement in validation accuracy (left panel)  without changing the validation cross-entropy loss (right panel, equation 1). The black vertical bars cover the 1st quartile to 3rd quartile of best validation accuracies or losses over model-ensemble members with a height of the bars representing the median.} 
\label{fig:ip_acc}
\end{figure}

\section{Dependence of Ensemble Performance on Inputs}
We gradually mask different types of inputs to record the change in model-ensemble performance. We train 3 different sets of model-ensemble with (1) Only magnetogram sequences, (2) Magnetogram sequences + Upstream properties, (3) Magnetogram sequences + Upstream properties + Wind/waves time-frequency image + X-ray time-series + Electron time-series.
We don't change the model architecture and instead mask the unused inputs with zeros. On the respective validation sets, we find clear improvement for the addition of each input (Figure~\ref{fig:ip_acc}). This should be noted that Wind/waves time-frequency data is not available in real-time and thus not used in the operational model. However, our results show the benefit of having that data in real-time.

\section{Design of Test set}
\begin{figure}[t!]
\hspace{-0.0\textwidth}
\includegraphics[width=1.0\linewidth]{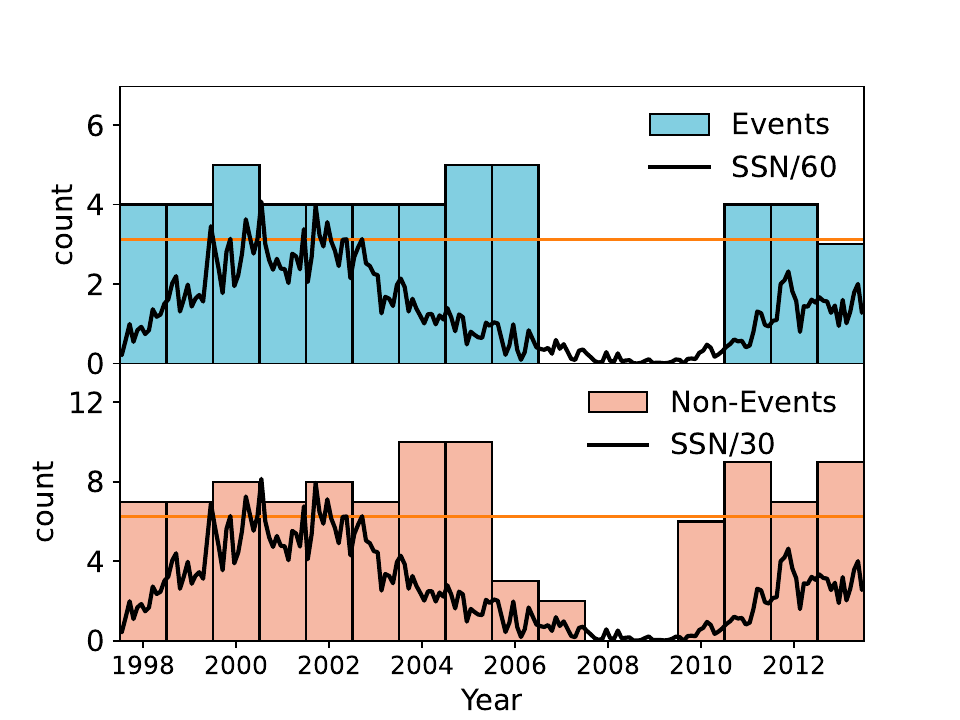}
\vspace{-0.08\textwidth}
\caption{Yearly histogram of the events (top panel; depicted in sky blue) and non-events (bottom panel; depicted in pink) depicted in blue. The horizontal orange line represents the event count per bin if the events or non-events were uniformly distributed across the bins. Smoothed monthly sunspot number cycle is shown with the black curve. It can be observed that event/non-event counts are not modulated by the sunspot number cycle for providing an unbiased estimate of model performance on the test set.}

\label{fig:test}
\end{figure}

We design a test set that represents all the flare classes and is not affected by the solar cycle phase in event selection (Figure~\ref{fig:test}). We select 5 events and 10 non-events per flare bin for all the 10 bins and also ensure that a year is dropped from random selection when it appears more than $\frac{N_{events}*N_{bins}}{N_{years}}$ times. 
We also ensure that the test set events are well separated in time from training and validation sets. This test set allows us to provide an unbiased estimate of model-ensemble performance.

\section{Probability calibration and reliability}

The `sigmoid' activation function ($sigmoid(x) = \frac{1}{1+e^{-x}}$) at the end of the classification model produces a number $p$, between 0 and 1. For $p$ to represent true probability of SEP occurrence, the relation: $f\approx p$ should hold where $f$ is the frequency of events around probability $p$. When we plot $p$ vs $f$ dividing all the predictions into probability bins we produce a reliability diagram where a diagonal line shows perfect reliability. When we generate this reliability diagram for training we don't find a perfect overlay of the ensemble on $y=x$ line. This implies the requirement of probability calibration. We use the Bayesian Binning Quantile (BBQ) model \cite{10.5555/2888116.2888120} to calibrate the CNN outcomes. This method gets rid of the disadvantages of other calibration techniques such as Platt Scaling and simple histogram binning \cite{10.5555/3305381.3305518}. Platt scaling suffers from the sigmoidal shape of the parametric model and simple histogram binning, a non-parametric model, suffers from the subjectivity in choosing the number of bins.
The BBQ model instead combines several equal-frequency histogram binning models with weights evaluated using Bayesian priors derived from a beta distribution.

\begin{figure}[htbp]
\hspace{-0.08\textwidth}
\includegraphics[width=1.1\textwidth]{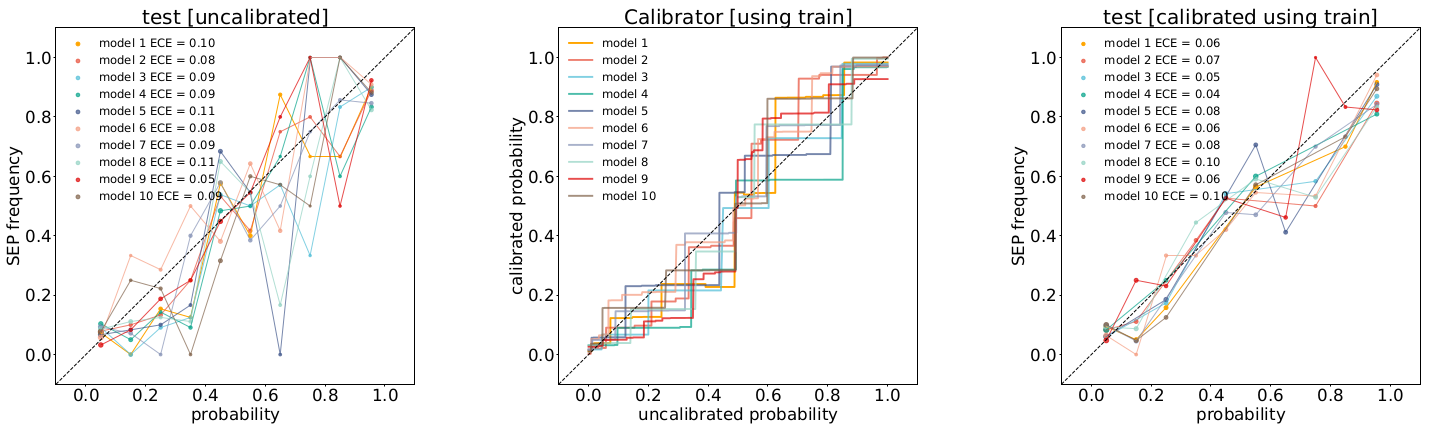}
\caption{Improvement in reliability post-calibration on unseen data. \textit{Left panel} shows uncalibrated probability (CNN outcome) vs. SEP event frequency for all the members of model-ensemble on test set with repective expected calibration errors as legends. The black digonal line represents perfect reliability. \textit{Middle panel} shows the calibrators, for all the ensemble members, that convert CNN outcome to true probability. These are generated by applying Bayesian Binning Quantile method on training set for respective ensemble members. \textit{Right panel} shows calibrated probability (using calibrators from training set) vs. SEP event frequency for all the members of model-ensemble on test set. }
\label{fig:bbq_results}
\end{figure}

\begin{linenomath*}
    \begin{equation}
        P = \sum_{i}Score_i\times P_i(NN)
    \end{equation}
\end{linenomath*}

where, $NN$ stands for the uncalibrated neural network outcome, $P_i(NN)$ represents the calibrated probability using binning model $i$ and $Score_i$ represents corresponding Bayesian score proportional $\prod_{b=1}^{B} \frac{\Gamma(N'/B)\Gamma(m_b+\alpha_b)\Gamma(n_b+\beta_b)}{\Gamma(N_b + N'/B)\Gamma(\alpha_b)\Gamma(\beta_b)}$.$B$ stands for number of bins in $i^{th}$binning model. $i$ runs over all possible histogram binning models. Different parameters shown in the Bayesian score expression are discussed in detail in \cite{10.5555/2888116.2888120}.  We evaluated the BBQ calibration curves on the training set and evaluate the effect of calibration on unseen data provided by the respective validation set and common test set. We quantify the deviation from the perfect reliability line with a measure called `Expected Calibration Error (ECE)' given by:

\begin{linenomath*}
    \begin{equation}
        ECE = \frac{1}{N}\sum_i n_i|f_i - <p_i>|
    \end{equation}
\end{linenomath*}
where $i$ runs over all the probability bins, $N$ is the total number instances, $n_i$ is the number of instances in bin $i$, $<p_i>$ represents the average `event' probability in bin $i$ and $f_i$ represents the frequency of events in bin $i$.
We use 10 bins for the evaluation of ECE and find improvement in ECE for 8 out of 10 models after calibration (Figure~\ref{fig:bbq_results}; median ECE changes from 0.09 to 0.065 after calibration).

\section{Performance metrics}

\subsection{Probabilistic metrics}
We emphasize that our objective here is not to do just a binary classification. Instead, we focus on estimating true probabilities of SEP occurrence with associated uncertainties (using a model-ensemble).   Our aim is to provide flexibility for the users of our forecast to determine their own acceptable level of risk, rather than imposing a threshold of detection that optimizes an arbitrary binary classification metric. The probabilistic metrics are described below-

\begin{enumerate}
    \item Brier Score (BS): $\frac{1}{N}\sum_{i=1}^{N}(t_i - p_i)^2$ while $t_i$ is the ground-truth binary label and $p_i$ is predicted `event' probability for instance $i$.
    \item  $ECE = \frac{1}{N}\sum_i n_i|f_i - <p_i>|$ described in section 7.
\end{enumerate}

\subsection{Binary classification metrics}
We evaluate the following performance metrics requiring a probability threshold and skill scores on the test set post-calibration for each ensemble member as well as model-median for comparison with \cite{bain21}:
\begin{enumerate}
    \item Probability of Detection (PoD): $\frac{TP}{TP + FN}$
    \item False Positive Rate (FPR): $\frac{FP}{TN + FP}$
    \item Accuracy (ACC): $\frac{TP + TN}{TP + TN + FP + FN}$
    \item True Skill Score (TSS): $\frac{TP}{TP + FN} + \frac{TN}{TN + FP} -1$
    \item Heidke Skill Score (HSS): $\frac{2(TP\times TN - FP\times FN)}{(TP + FN)(TN + FN) + (TN + FP)(TP + FP)}$
    \item Area Under receiver operating characteristic Curve (AUC): for this we plot PoD vs. FPR for different probabilistic decision thresholds used by the model to classify an event or non-event prediction. The Area Under the Curve generates the `AUC' metric which gives an indication of how well the model performs in relation to a forecast of random chance.
\end{enumerate}

where TP is the number of true positives i.e. the number of instances for which the model gave a correct event forecast, and FP is the number of false positives, i.e. the number of instances for which the model incorrectly forecast an event when no event occurred, TN is the number of true negatives i.e. the number of instances for which the model correctly forecast no proton event (i.e. non-event) associated with the flare, and FN is the number of instances for which the model failed to forecast an event which did occur.

\section{Model-ensemble Performance on test set}
\begin{figure}[t!]
\hspace{-0.1\textwidth}
\includegraphics[width=1.2\linewidth]{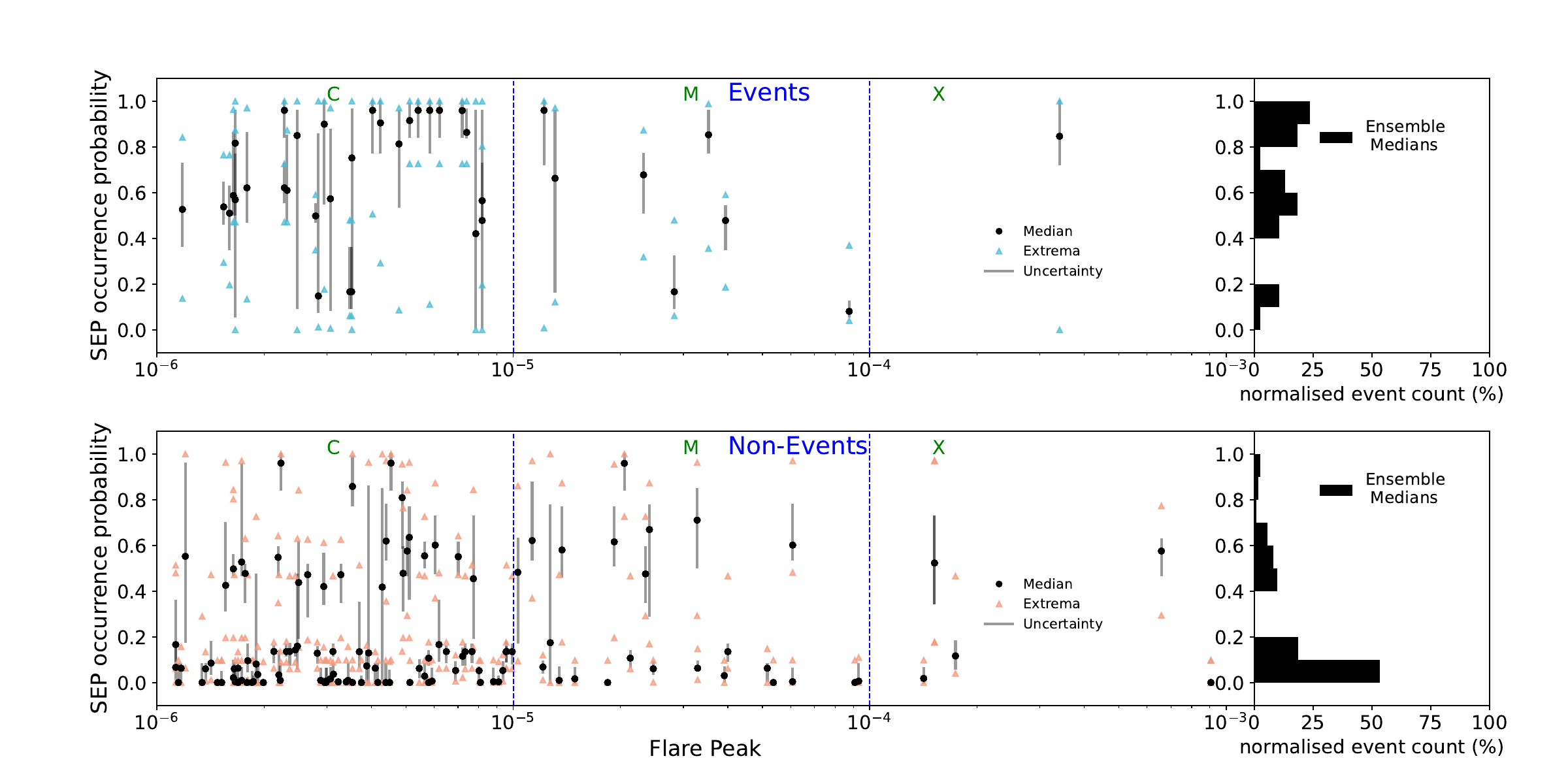}
\caption{Model-ensemble inference on each event and non-event of the test set. The top panel shows the inference on events and the bottom panel depicts the inference on non-events. X-axis marks the associated flare peak for each event/non-event and Y-axis represents the calibrated SEP occurrence probability predicted by the ensemble of 10 models. Model-ensemble inference extrema are depicted by colored (sky blue for events; pink for non-events) triangles and the black vertical lines show the uncertainty in SEP occurrence probability. Ensemble medians are depicted by a large black dot. The marginal histograms (in black) represent the distribution of ensemble-median probabilities with 10 bins. These provide important information to the user to appropriately choose the probability threshold.}
\label{fig:class_pred}
\end{figure}

We plot the calibrated inferred probabilities of SEP event occurrence for each event and non-event in the test set sorted by peak flare strength (Figure~\ref{fig:class_pred}). We evaluate model-ensemble median for each inference to estimate different performance metrics and also depict the interquartile range for ensemble inference as the prediction uncertainty (Figure~\ref{fig:class_pred}). In general, we find smaller uncertainty in ensemble inference when the medians are close to 0 for non-events and close to 1 for events. The uncertainty increases when medians move to the false prediction regime. This provides an understanding of whether an event is well represented in the training set and how actionable a prediction is. 

We estimate different skill scores (as described in the previous section) both on each ensemble member inference and ensemble median. The estimated probabilistic and binary classification metrics are shown in Table~\ref{skill_scores}. For binary classification metrics we use a commonly used probability threshold of 0.5 \cite{bain21}.

 \begin{table}[]
  \caption{Skill scores of model-ensemble on test set}
 \hspace{-0.05\textwidth}
\begin{tabular}{ccccccccccccc}
                                           & Skill Scores    & Ensemble-Median & M1 & M2 & M3 & M4 & M5 & M6 & M7 & M8 & M9 & M10 \\ \hline \hline
\multicolumn{1}{l|}{\multirow{2}{*}{\rotatebox[origin=c]{90}{Prob.}}}  & BS  &  0.14 & 0.15 & 0.15 & 0.14 & 0.15 & 0.14 & 0.15 & 0.15 & 0.16 & 0.15 & 0.13                   \\ 
\multicolumn{1}{l|}{}                     & ECE &  0.07 & 0.06 & 0.07 & 0.05 & 0.04 & 0.08 & 0.06 & 0.08 & 0.1 & 0.06 & 0.1    \\ \hline
\multicolumn{1}{l|}{\multirow{6}{*}{\rotatebox[origin=c]{90}{Binary}}} & PoD &  0.83 & 0.85 & 0.55 & 0.57 & 0.81 & 0.83 & 0.64 & 0.66 & 0.79 & 0.57 & 0.85                    \\ 
\multicolumn{1}{l|}{}                     & ACC &  0.81 & 0.79 & 0.78 & 0.79 & 0.8 & 0.81 & 0.77 & 0.77 & 0.77 & 0.77 & 0.83                    \\
\multicolumn{1}{l|}{}                     & TSS &  0.63 & 0.6 & 0.46 & 0.48 & 0.6 & 0.63 & 0.49 & 0.48 & 0.55 & 0.45 & 0.67                  \\ 
\multicolumn{1}{l|}{}                     & HSS &  0.6 & 0.57 & 0.49 & 0.51 & 0.57 & 0.6 & 0.49 & 0.49 & 0.53 & 0.46 & 0.65                    \\ 
\multicolumn{1}{l|}{}                     & FPR & 0.2 & 0.25 & 0.09 & 0.09 & 0.21 & 0.2 & 0.15 & 0.18 & 0.24 & 0.13 & 0.18                     \\
\multicolumn{1}{l|}{}                     & AUC & 0.87 & 0.84 & 0.85 & 0.86 & 0.84 & 0.86 & 0.84 & 0.84 & 0.83 & 0.85 & 0.87                     \\ 
\end{tabular}
\label{skill_scores}
\end{table}

\section{Model validation on SHINE Challenge 2022 dataset}
\begin{figure}[!htbp]
\hspace{-0.02\textwidth}
\includegraphics[width=1.1\textwidth]{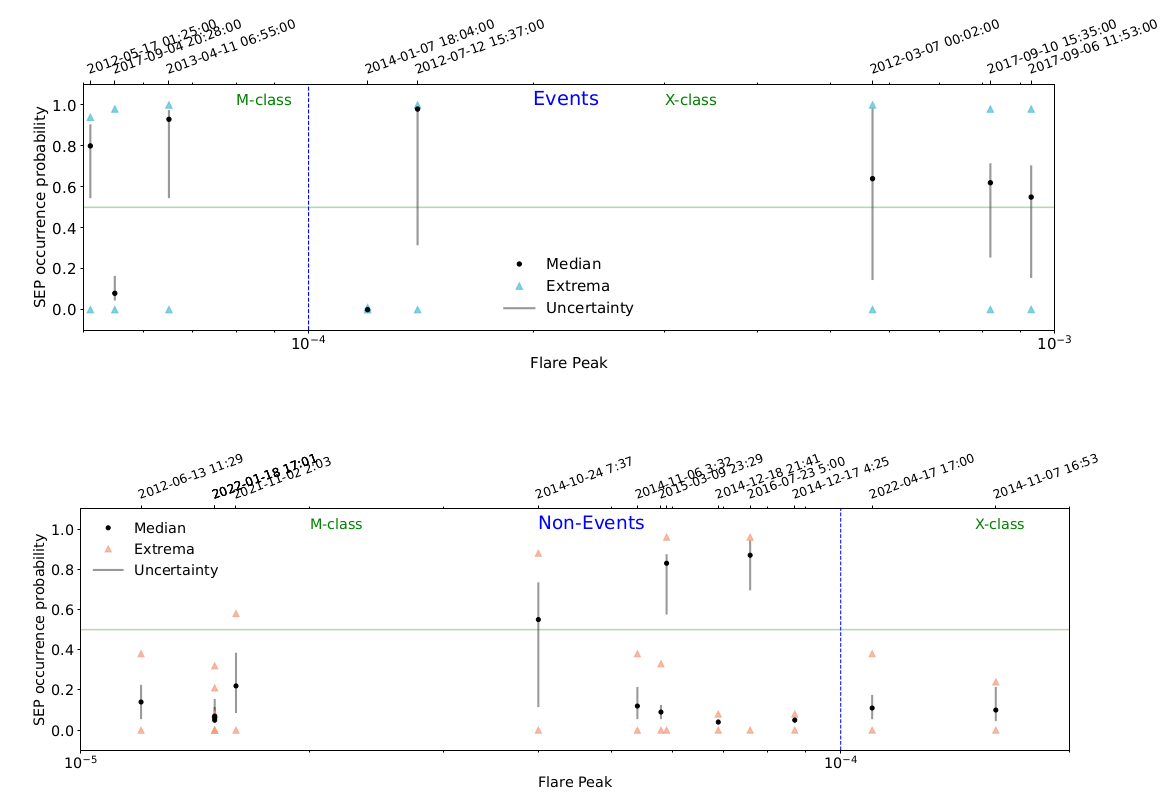}
\caption{Performance of MEMPSEP on SHINE 2022 SEP model validation challenge events (top panel) and non-events (bottom panel). Model-ensemble inference extrema are depicted by a colored (sky blue for events; pink for non-events) triangles and the gray vertical lines show the uncertainty in SEP occurrence probability. Ensemble-medians are
depicted by the large black dots. X-axis depicts associated M, X-class flare strengths. Flare onset times associated with events and non-events are shown on the upper x-axis of each panel.}
\label{fig:shine_validation}
\end{figure}

We validated MEMPSEP during Solar Heliospheric and INterplanetary Environment (SHINE) 2022 SEP model validation Challenge against 14 non-events between 2012-2022 and 8 events during 2012-2017 associated with M and X-class flares. We provided our model-ensemble calibrated probabilities for each of those as depicted in Figure~\ref{fig:shine_validation}. We received a Brier score of 0.2 based on model-median prediction. In Binary setting with a probability threshold of 0.5 MEMPSEP correctly 6 out of 8 events and 11 out of 14 non-events. As found in the test-set we found higher uncertainty in forecast for SHINE events as compared to non-events.

\section{Discussion and Conclusion}
We describe a new SEP predictive model, MEMPSEP, which is a CNN model-ensemble that ingests both remote-sensing and in-situ inputs and predicts SEP occurrence probability (i.e. probability of proton flux crossing 5 pfu in the energy band $>$10 MeV within 6 hours from flare onset) and physical properties. 
This paper focuses on forecasting the true probability of SEP occurrence and associated uncertainty with MEMPSEP. The ensemble members are built utilizing the large class imbalance between events and non-events.

We calibrate the outcome of each model to represent true probability and find improvements ($\sim1-5\%$ over the ensemble members) in model reliability post-calibration. We also achieve a Brier score of 0.14 for our test set and 0.2 for SHINE Challenge 2022 events, non-events based on ensemble-median probability. We emphasize that having both true probability and associated uncertainty will provide the users enough flexibility to tune the model-ensemble outcome to their forecasting priority.

We would like to highlight that even though the events and non-events in our dataset are tied to flares, we do not use the flare peak as input and instead utilize input data only until the flare onset. It can thus be tested as a future work if MEMPSEP can ingest data during the flare quiet period further back in time from flare onset to predict SEP occurrence.

Also as a future follow-up, we quantify the effect of the number of ensemble members on uncertainty and investigate the conditions of its convergence.

We have validated MEMPSEP with the campaign events/non-events of SHINE and provided the results to the heliospheric and space weather community. We continue to participate in SHINE validations and present MEMPSEP to the broader community, namely, International Space WEather Actions Teams (ISWAT), CCMC, and relevant organizations. 
Future tasks also include adding additional input layers to MEMPSEP (e.g., \citeA{Elliott2022}) and performing comparative studies with other operational SEP prediction models to determine the scope of improvements. 

\newpage
\section{Open Research}
The dataset used in this paper comes from public repositories (e.g.  Heliospheric Event Knowledgebase (HEK), JSOC) and are rigorously described in MEMPSEP-III. The codebase developed for this work can be accessed from \url{https://github.com/subhamoysgit/MEMPSEP} \cite{MEMPSEPgithub}.

\acknowledgments
This work is mainly supported by SWO2R grant 80NSSC20K0290. Partial support for MAD, KM, and SH comes from SW2OR 80NSSC21K0027, NASA LWS grants 80NSSC19K0079, 80NSSC21K1307, and 80NSSC20K1815.

\end{document}